\documentclass{amsart}
\usepackage[dvips]{graphicx}
\usepackage{latexsym}
\usepackage{amsmath}
\usepackage{lipsum}
\usepackage[varg]{txfonts}
\usepackage{here}
\usepackage{setspace}
\begin{document}
\begin{center}
\begin{spacing}{2}
\begin{Huge}
Experiments Validating the Effectiveness of Multi-point Wireless Energy Transmission with Carrier Shift Diversity
\end{Huge}
\end{spacing}
\end{center}

\begin{large}
\begin{center}
Daiki Maehara$^{1}$, {Gia Khanh} Tran$^{1}$, Kei Sakaguchi$^{1}$, Kiyomichi Araki$^{1}$ \\ \and Minoru Furukawa$^{2}$ \\ 
\begin{spacing}{2}
\end{spacing}
$^{1}$Tokyo Institute of Technology, Tokyo, Japan \\
Email: {maehara@mobile.ee.titech.ac.jp} \\
$^{2}$Nihon Dengyo Kosaku Co., Ltd, Saitama, Japan
\end{center}
\end{large}
\begin{spacing}{2}
\end{spacing}
\begin{flushright}
DRAFT: July 2014
\end{flushright}

\begin{abstract}
This paper presents a method to seamlessly extend the coverage of energy supply field for wireless sensor networks in order to free sensors from wires and batteries, where the multi-point scheme is employed to overcome path-loss attenuation, while the carrier shift diversity is introduced to mitigate the effect of interference between multiple wave sources. As we focus on the energy transmission part, sensor or communication schemes are out of scope of this paper. To verify the effectiveness of the proposed wireless energy transmission, this paper conducts indoor experiments in which we compare the power distribution and the coverage performance of different energy transmission schemes including conventional single-point, simple multi-point and our proposed multi-point scheme. To easily observe the effect of the standing-wave caused by multipath and interference between multiple wave sources, 3D measurements are performed in an empty room. The results of our experiments together with those of a simulation that assumes a similar antenna setting in free space environment show that the coverage of single-point and multi-point wireless energy transmission without carrier shift diversity are limited by path-loss, standing-wave created by multipath and interference between multiple wave sources. On the other hand, the proposed scheme can overcome power attenuation due to the path-loss as well as the effect of standing-wave created by multipath and interference between multiple wave sources. 
\end{abstract}

\section{Introduction}
Recently, Wireless Sensor Networks (WSNs) are used for many applications, e.g. health monitoring of buildings, factory automation, and energy management systems. For example, in Building Energy Management System (BEMS), numerous sensors are deployed in fields such as office room to observe environmental information, e.g. temperature, brightness, human detection, and so on. The information are gathered via wireless links and employed for saving power consumption. However limited life-time of sensors has long been an issue in WSNs. Conventionally, sensor's energy is supplied by electrical plug, battery or environmental energy, which respectively have the following disadvantages: limitation of installation placement, requirement of battery replacement, or lack of stability. To deal with the problem, we propose to introduce wireless energy transmission to WSNs, in which sensors can be released from both wires and batteries and be stably supplied with energy. 

Wireless energy transmission schemes are categorized into three types, i.e. radio wave emission, resonant coupling and inductive coupling \cite{book1}. In the radio wave emission method, energy is collected at the receiver (Rx) using rectenna (rectifying antenna) to receive and convert radio wave into direct current. Compared with the other schemes, long range transmission can be realized by increasing transmit power or antenna gain. Space Solar Power Satellite (SSPS) is one of the examples of this scheme. In WSNs, numerous ubiquitous sensors are distributed in indoor environments. In order to supply energy to all of them, the radio wave emission is employed in this paper.

So far, radio wave emission technology has been mainly used for Radio Frequency IDentification (RFID) systems. However, RFID systems have not been designed for the wide-area coverage targeted in this paper. Additionally, in passive RFID systems in which sensors do not have external battery, radio wave is emitted only when Reader/Writers (R/Ws) request tag data. Consequently, sensors cannot transmit the data on their own initiative.  Furthermore, in a single transmitter (Tx) system, the coverage of energy supply field is restricted by maximum transmit power limited by the radio regulation. To extend the coverage, multiple transmitters can be introduced to the systems. This, however, results in the collision between multiple transmitters. To avoid the collision while complying with the regulation, a collision avoidance scheme among multiple R/Ws, e.g. Time Division Multiple Access (TDMA), should be employed \cite{TDMA}. However, employing TDMA results in decreasing the time efficiency in terms of energy charging and increasing the system complexity in proportion to the number of introduced R/Ws. On the other hand, if multiple transmitters simultaneously perform energy transmission, destructive interference between multiple wave sources results in deadspots where sensors cannot be activated. This scheme is called simple multi-point in this paper.

\begin{figure} [t]
\centering
\includegraphics[width=12cm]{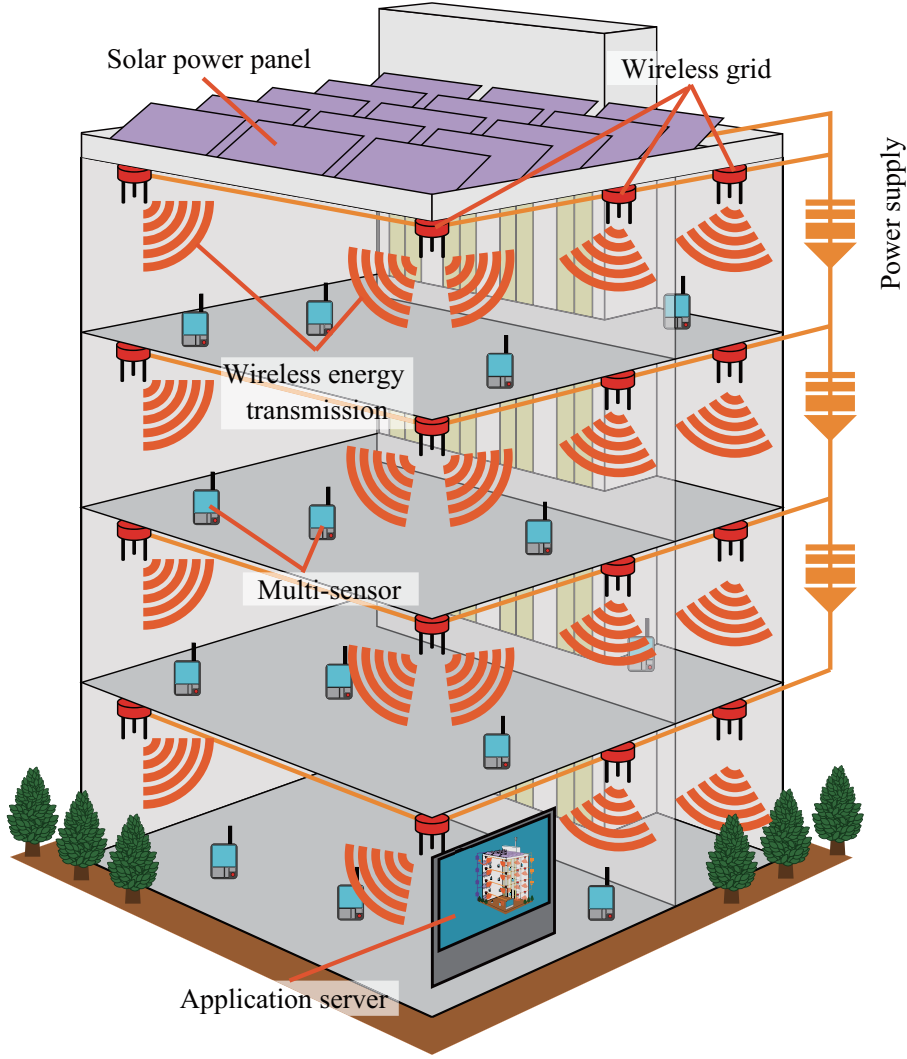} 
\caption{One of the examples of wireless grid application.}
\label{fig:WG} 
\end{figure}

To tackle these problems, we propose a system called the wireless grid to seamlessly supply wireless energy to sensors as shown in Fig.~\ref{fig:WG} \cite{Paper0}\cite{Paper1}. Grid nodes, located at ceilings or integrated in fluorescent lamps, continuously supply wireless energy to sensors to charge their rechargeable batteries, e.g. capacitor. This energy is utilized at the sensors for performing both environmental sensing and communication with the grid nodes. The wireless grid can expand the energy coverage due to our proposal of multi-point wireless energy transmission with carrier shift diversity. Owing to the introduced carrier shift diversity, which was originally proposed for data communication \cite{phaseshift1}\cite{phaseshift2}, artificial fading can be created to cancel out deadspots by time averaging. By this way, wireless grid can seamlessly extend the coverage and supply energy to all sensors distributed in indoor environments. 

Several companies and researchers have attempted introducing wireless energy transmission into wireless sensor networks by employing single-point RFID systems or harvesting the ambient RF power \cite{PC}-\cite{TU}. In these papers, \cite{GT} proposed to transmit the special waveform to improve the coverage. \cite{CU} employed two orthogonal polarization and narrowband frequency modulation to improve the uniformity of the power density in a metallic over-moded waveguide cavity. However, the single-point scheme is difficult to improve the coverage due to path-loss and multi-path. In addition, harvesting the ambient power depends on the environment. On the other hand, \cite{Room}\cite{WU} employed multi-transmitter or multi-antenna systems to improve the received power of certain sensors with fixed location by phase control on each antenna. 

Different from these researches, this paper realizes seamless coverage of energy supply field to activate all sensors distributed in the indoor environments by the multi-point wireless energy transmission with carrier shift diversity. Furthermore, as a proof of concept, we conduct indoor experiments, partially presented in \cite{ISSSE}, to verify the effectiveness of the proposed scheme. In our experiments, we compare the received power distribution as well as the coverage in a single-point scheme, a conventional multi-point scheme, and the proposed multi-point scheme in the same environment. The experimental results show that the proposed scheme can overcome power attenuation due to the path-loss as well as the effect of standing-wave created by multipath and interference between multiple wave sources. In particular, the maximum available value of required power to maintain 100\% coverage in the proposed scheme is improved by 18 dB compared with that in the single-point case.

For the rest of this paper, Sec.~\ref{sec:2} gives theoretical support for the benefits of the proposed multi-point wireless energy transmission. The theory is validated by both experiments in real indoor environment and simulations assuming free space in Sec.~\ref{sec:3}. Finally, Sec.~\ref{sec:4} concludes this paper.   

\clearpage
\section{Multi-point Wireless Energy Transmission \\with Carrier Shift Diversity}
\label{sec:2}

\subsection{Spectrum Allocation for Wireless Energy Transmission}
\begin{figure} [t]
\centering
\includegraphics[width=12cm]{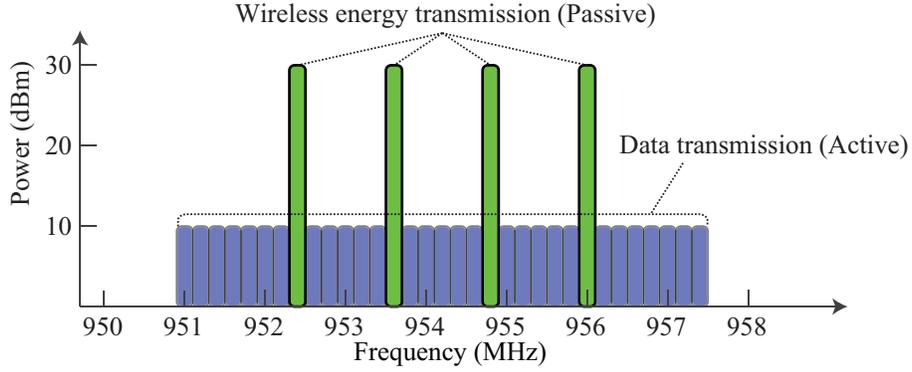} 
\caption{The available frequency spectrum in 950 MHz band. Green for wireless energy transmission: maximum transmit power is 30 dBm and bandwidth is 200 kHz. Blue for data communication: maximum transmit power is 10 dBm and bandwidth is 200 kHz.}
\label{fig:950} 
\end{figure}

Figure~\ref{fig:950} shows the spectrum mask of 950 MHz band which is available for wireless energy transmission including UHF RFID systems. According to Japanese regulation \cite{Reg1},  the maximum transmit power and the maximum Equivalent Isotropically Radiated Power (EIRP) are respectively limited to 30 dBm and 36 dBm over an energy transmission channel of 200 kHz bandwidth. There are four non-LBT (Listen Before Talk) channels where transmission without carrier sensing is allowed. In our proposed wireless grid, these four channels are all continuously and simultaneously employed for wireless energy transmission while the other channels are used for data communication. The 950 MHz band in Japan has been reallocated for LTE mobile systems while the 920 MHz band is now available for RFID systems with wireless energy transmission. It is noted that these four non-LBT channels are shifted to the 920 MHz band in the same format.

\begin{figure} [h]
\centering
\includegraphics[width=12cm]{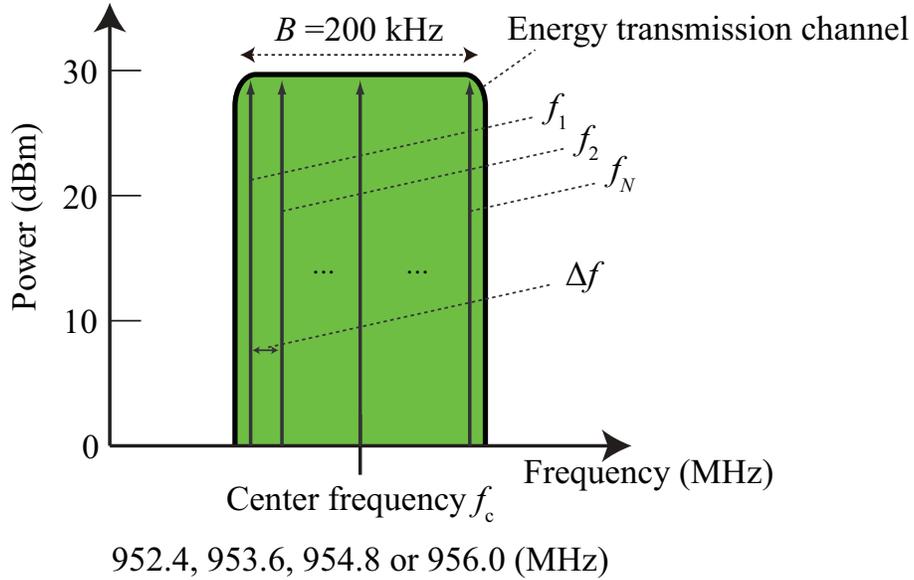} 
\caption{Frequency allocation for carrier shift diversity.}
\label{fig:CSD} 
\end{figure}

In the carrier shift diversity, the center frequency of the carrier of each energy transmission point is slightly shifted with a predefined amount to create artificial fading. The time-varying fading shuffles the interference pattern between multiple wave sources with suitable selection of carrier offsets. In multi-point wireless energy transmission with carrier shift diversity, the available frequency bandwidth of 200 kHz is divided into $N$ orthogonal subcarriers and these subcarriers are respectively allocated to $N$ different wireless energy transmission points. 
The carrier frequency of the $n\mathrm{-th}$ Tx can be defined as 
\begin{eqnarray}
f_{n}  = f_{\mathrm{c}}-\frac{B}{2}+\frac{\left(2n -1 \right)B}{2N},\ \ \ \ (n=1,\ldots ,N) ,
\label{equ:FD}
\end{eqnarray}
where $f_{c}$ and $B$ respectively denote the center frequency and the channel bandwidth as shown in Fig.~\ref{fig:CSD}.

\begin{figure} [h]
\centering
\includegraphics[width=12cm]{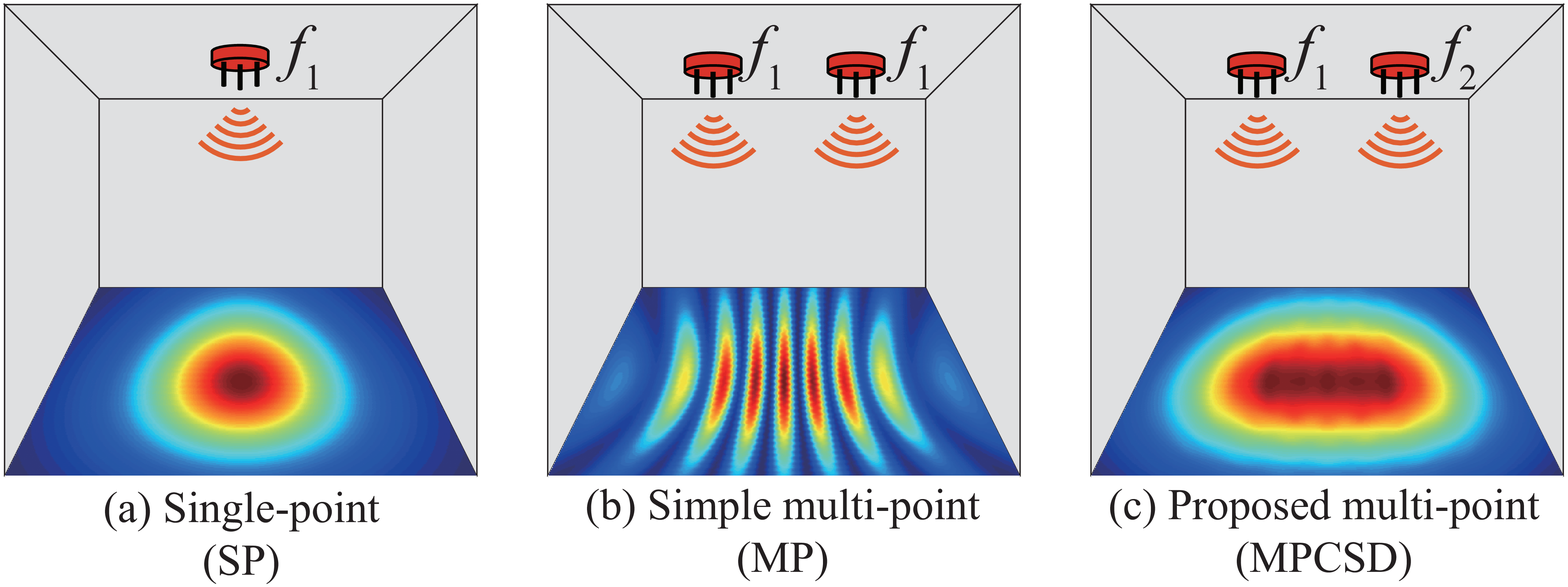} 
\caption{Power distribution of wireless energy transmission schemes.}
\label{fig:MPC} 
\end{figure}

\subsection{Concept of Multi-point Wireless Energy Transmission}
 In this paper, the conventional single-point, simple multi-point and our proposed multi-point are abbreviated as SP, MP and MPCSD respectively henceforth. Figure~\ref{fig:MPC} shows one example of power distribution when employing these three schemes.

In Fig.~\ref{fig:MPC}~(a),  the coverage of SP is limited by the maximum transmit power defined by the radio regulation. Furthermore, in real environments,  the effect of standing-wave created by multipath becomes more remarkable in the case when the power difference between the direct and reflected waves is small. This effect will also occur due to reflections from floor and ceiling when horizontal polarized waves are concerned \cite{book2}. Due to this effect,  it is difficult to provide seamless energy transmission to sensor nodes since the sensor nodes might be located at the deadspots. For example, \cite{Mizuho} reported that R/Ws in real RFID systems cannot read IC tags even located at a distance shorter than the designed coverage of the product.

To solve the coverage limitation due to path-loss in SP, additional transmission points can be introduced as shown in Figs.~\ref{fig:MPC}~(b) and~(c) to enhance the area of energy supply field. However, merely increasing the number of energy transmission points does not solve the limitation of the coverage as shown in Fig.~\ref{fig:MPC}~(b). The effect of interference between multiple wave sources can be avoided by applying a TDMA scheme to MP. However, the time efficiency of energy supply decreases and the complexity of the system increases in proportion to the numbers of transmitters. To deal with the problem, we propose to apply carrier shift diversity to MP.

By using the carrier shift diversity,  the destructive interference can be significantly alleviated as shown in Fig.~\ref{fig:MPC} (c) while energy can be simultaneously supplied by multiple transmitters. In other words, the proposed method can realize a seamless coverage extension without reducing the time efficiency of energy supply compared to that in MP.

\subsection{Theoretical Discussion}
\begin{figure} [t]
\centering
\includegraphics[width=12cm]{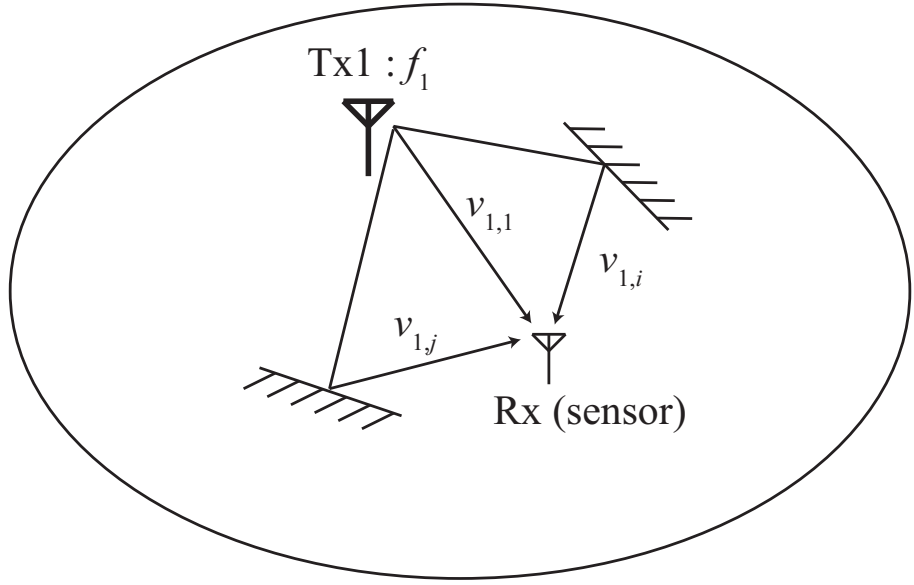} 
\caption{System model of SP.}
\label{fig:SP_sys} 
\end{figure}

Assuming a system model for SP as shown in Fig.~\ref{fig:SP_sys}, the received voltage at the sensor node from the $i$-th path ($i=1, 2, \ldots, I$) is represented as
\begin{eqnarray}
v_{1,i}\left(t\right) = \sqrt{2}V_{1,i}\left(r_{1}\right) \cos \left[\omega_{1}t+\theta_{1,i}\left(r_{1}\right)\right] , 
\end{eqnarray}
where $r_{1}$ is the distance between the Tx and the Rx, $V_{1,i}\left(r_{1}\right)$ is the DC voltage obtained from the $i$-th path,  $\omega_{1}=2 \pi f_{1}$ denotes the angular frequency with carrier frequency of $f_{1}$,  and $\theta_{1,i}\left(r_{1}\right)$ is an initial phase. 
The voltage obtained from all of paths becomes
\begin{eqnarray}
v_{1}\left(t\right) = \sum_{i=1}^{I}{v_{1,i}\left(t\right)}.
\end{eqnarray}
According to \cite{Rec}, a single-shunt rectenna converts RF signal into DC power with 100\% efficiency by using a perfect matching network and an ideal diode with lossless property. Therefore, the average received power can be described as
\begin{eqnarray}
P_{1} &=& \frac{1}{R_\mathrm{out}}\overline{{v_{1}}^{2}\left(t\right)}  \nonumber\\
&=& \frac{1}{R_\mathrm{out}}\left( \sum_{i=1}^{I}V_{1,i}^2\left(r_{1}\right) \right. \nonumber \\
&+&\left. \sum_{i=1}^{I}\sum_{j \neq i}^{I}V_{1,i}\left(r_{1}\right)V_{1,j}\left(r_{1}\right) \cos \left[\theta_{1,i}\left(r_{1}\right)-\theta_{1,j}\left(r_{1}\right)\right]\right),  \nonumber \\
\label{equ:SP_power}
\end{eqnarray}
where $\bar{ \left[ \ \right] }$ and $R_{\mathrm{out}}$ respectively express the function of time averaging and the output load of sensor nodes.
This equation implies that the average received power in SP is affected by the effect of standing-wave created by multipath as shown in the second term of Eq.~(\ref{equ:SP_power}). In general, $V_{1,i}\left(r_{1}\right)$ is a function of the length of each path and its related reflection coefficients. According to Fresnel equations, reflection coefficient depends on polarization and incidence angle to the surface of obstacles. Therefore, the multipath components $v_{1,i}(t)$ with short path lengths or high reflection coefficient, are dominant in the received power  $P_{1}$.

\begin{figure} [t]
\centering
\includegraphics[width=12cm]{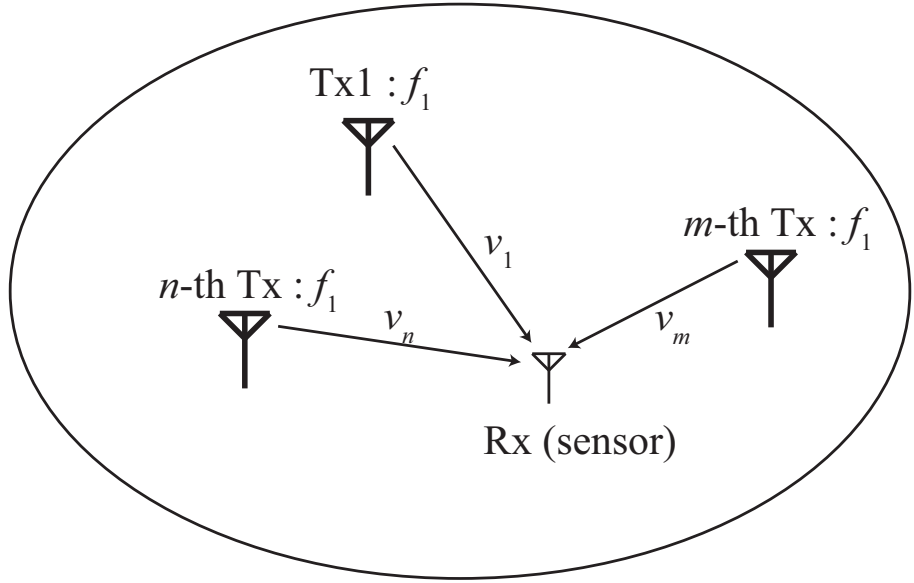} 
\caption{System model of MP.}
\label{fig:MPwo_sys} 
\end{figure}

\begin{figure} [h]
\centering
\includegraphics[width=12cm]{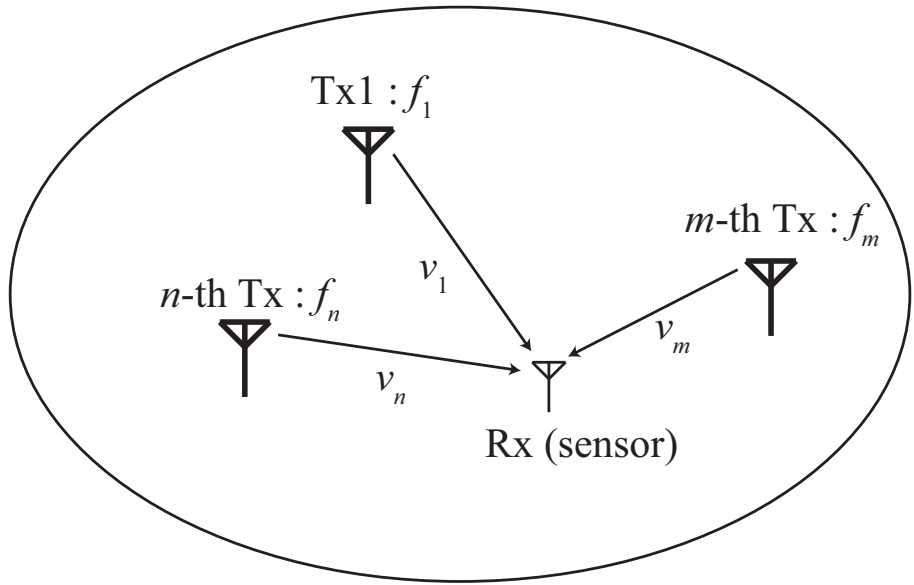} 
\caption{System model of MPCSD.}
\label{fig:MPw_sys} 
\end{figure}

Assuming a system model for MP and MPCSD as in Figs.~\ref{fig:MPwo_sys} and \ref{fig:MPw_sys} respectively, the received voltage at the sensor node obtained from the $n$-th Tx ($n=1,2, \ldots, N$) is represented as
\begin{eqnarray}
v_{n} \left( t \right)  = \sum_{i = 1}^{I} v_{n,i} \left( t \right).
\label{equ:single}
\end{eqnarray}
To simplify this equation, we can convert Eq. (\ref{equ:single}) to the following equation    
\begin{eqnarray}
v_{n} \left( t \right) = \sqrt{2}V_{n}\left(r_{n}\right)\cos \left[ \omega_{n} t + \theta_{n}\left(r_{n}\right) \right],
\label{equ:sig} 
\end{eqnarray}
where $r_{n}$ is the distance between the $n$-th Tx and the Rx,  $V_{n}\left(r_{n}\right)$ is the DC voltage obtained from the $n$-th Tx including the multipath effect,  $\omega_{n}=2 \pi f_{n}$ denotes the angular frequency with carrier frequency of $f_{n}$,  and $\theta_{n}\left(r_{n}\right)$ is an initial phase as a result of standing-wave created by multipath.
The received voltage obtained from all the transmitters becomes
\begin{eqnarray}
V \left( t \right)  = \sum_{n=1}^{N}v_{n} \left( t \right).
\end{eqnarray}
The average received power in the case of MP can be described as    
\begin{eqnarray}
P_{\mathrm{total}} &=& \frac{1}{R_\mathrm{out}}\overline{V^{2}\left(t\right)}  \nonumber\\
&=& \frac{1}{R_\mathrm{out}}\left(\sum_{n=1}^{N}V^2_{n}\left(r_{n}\right) \right. \nonumber \\
&+& \left. \sum_{n=1}^{N}\sum_{m \neq n}^{N}V_{n}\left(r_{n}\right)V_{m}\left(r_{m}\right) \cos \left[\theta_{n}\left(r_{n}\right)-\theta_{m}\left(r_{m}\right)\right]\right). \nonumber \\
\label{equ:wo}
\end{eqnarray}
On the other hand,  in the case of MPCSD, according to the effect of the carrier shift, the average received power is given as
\begin{eqnarray}
P_{\mathrm{total}}^{\mathrm{CS}} &=& \frac{1}{R_\mathrm{out}}\overline{V^{2}\left(t\right)} \nonumber \\
&=& \frac{1}{R_\mathrm{out}} \left( \sum_{n=1}^{N}V_{n}^2\left(r_{n}\right) + \sum_{n=1}^{N}\sum_{m \neq n}^{N} V_{n}\left(r_{n}\right)V_{m}\left(r_{m}\right) \right.  \nonumber \\
&\times& \left. \overline{ \cos \left[\left(\omega_{n}-\omega_{m}\right)t+\theta_{n}\left(r_{n}\right)-\theta_{m}\left(r_{m}\right)\right]} \right)  \nonumber \\
&=& \frac{1}{R_\mathrm{out}}\sum_{n=1}^{N}V^2_{n}\left(r_{n}\right)
\end{eqnarray}
Here, the received power is averaged over a period of $T=N/B$, which is the largest period of the artificial fading of MPCSD. In this equation, the artificial fading created by carrier shift diversity cancels out the second term of Eq.~(\ref{equ:wo})  (the effect of interference between multiple wave sources) by time averaging. Therefore, the average received power turns into the summation of the power from all transmitters. Although the received power obtained from the $n$-th transmitter $V^2_{n}/R_{\mathrm{out}}$ is independently affected by standing-wave caused by multipath, the probability that all of the received powers become deadspots is significantly small in comparison to the case of SP.  It is noted that the artificial fading period $T$ should be smaller than the minimum period of data transmission $T_{\mathrm{D}}$ to supply the average received power to the sensor IC. Therefore, $T_{\mathrm{D}}$ should satisfy the following inequality, 
\begin{eqnarray}
T_{\mathrm{D}} \geq  N/B = T.
\end{eqnarray} 
In practical systems, since the effect of interference between multiple wave sources is reduced in proportion to the distance between transmitters, about 20 subcarriers might be sufficient for indoor environment. In addition, since the available frequency bandwidth $B$ is 200 kHz, the duty cycle of sensor nodes should be more than 100$\mu$s which is sufficient to perform the transmission of sensing data. Therefore, the carrier shift diversity can be employed in WSNs.

\clearpage
\section{System Performance Evaluation}
\label{sec:3}
In this section, we aim to evaluate the concept of MPCSD in real environments. Therefore, we conduct experiments in indoor environment as shown in Fig.~\ref{fig:Ex_model}. As mentioned, the effect of standing-wave created by multipath has an influence to the energy coverage, so that we perform the measurement in a three dimensional space in order to observe this effect. In addition, for comparison with the ideal case of no multipath environment we conduct a simulation assuming a free-space environment. 

\subsection{Experiment Method}
\begin{figure}  [h]
\centering
\includegraphics[width=12cm]{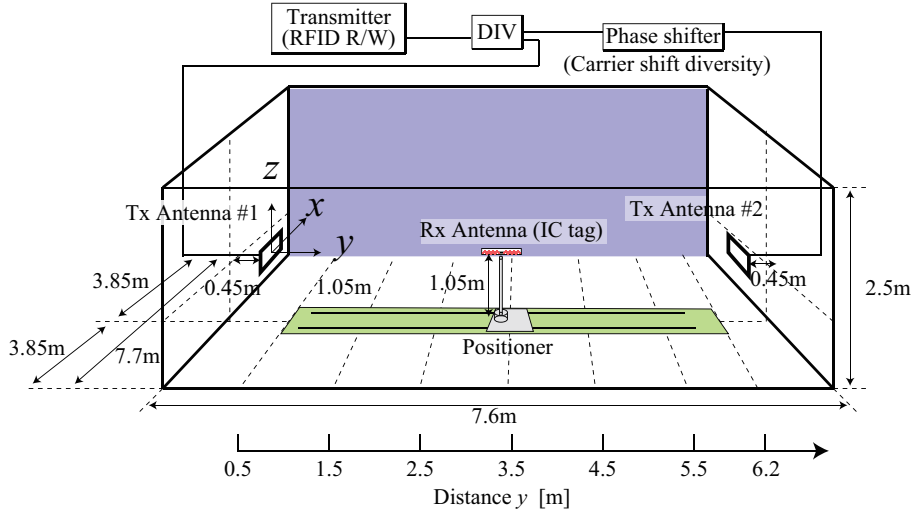} 
\caption{Experimental system for energy transmission.}
\label{fig:Ex_model} 
\end{figure}

\begin{figure}  [h]
\centering
\includegraphics[width=12cm]{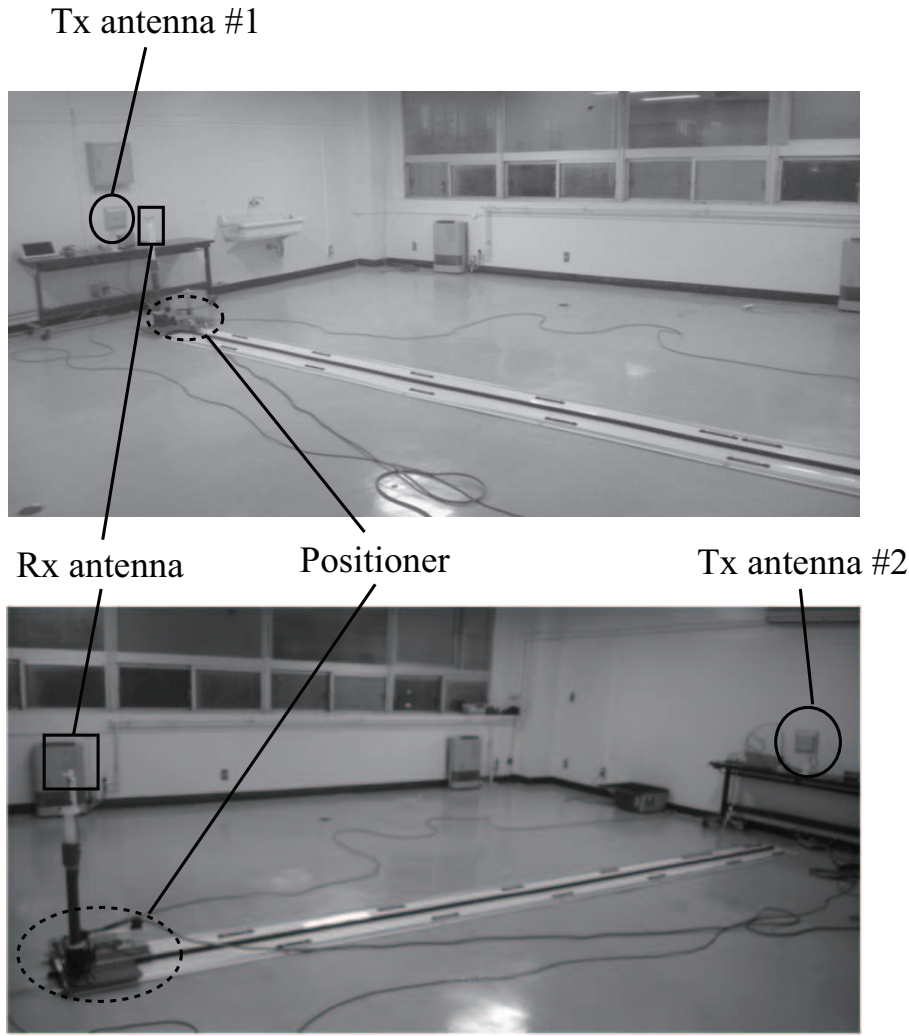} 
\caption{Experimental environment.}
\label{fig:Ex_model_pic} 
\end{figure}

\begin{figure}  [h]
\centering
\includegraphics[width=12cm]{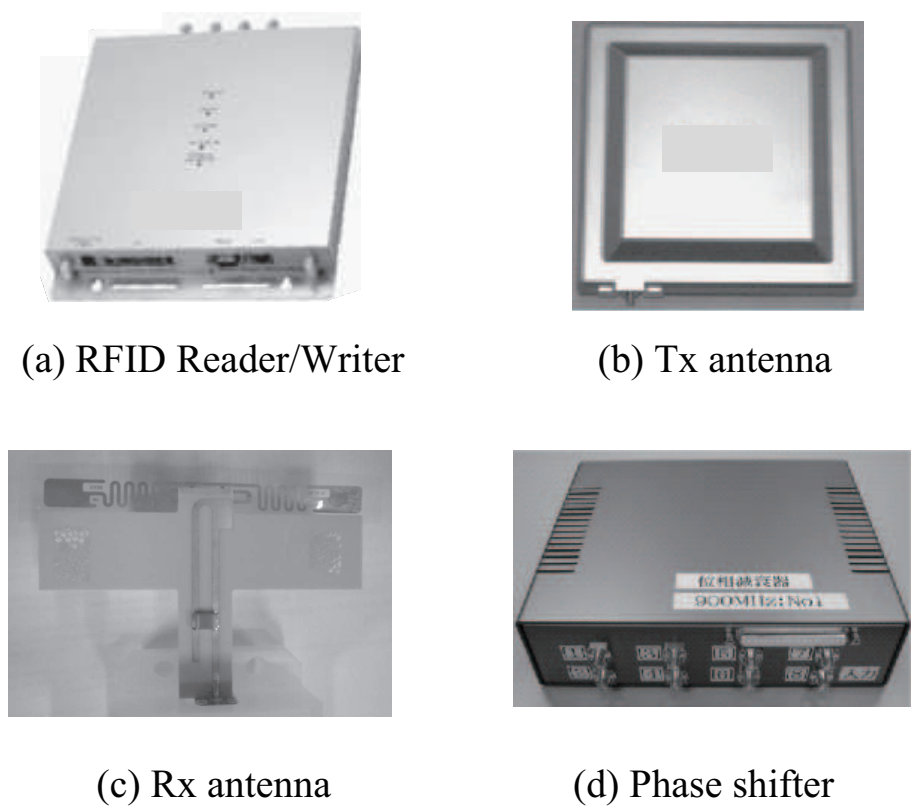} 
\caption{Experimental equipments.}
\label{fig:ins} 
\end{figure}

\begin{figure}  [h]
\centering
\includegraphics[width=12cm]{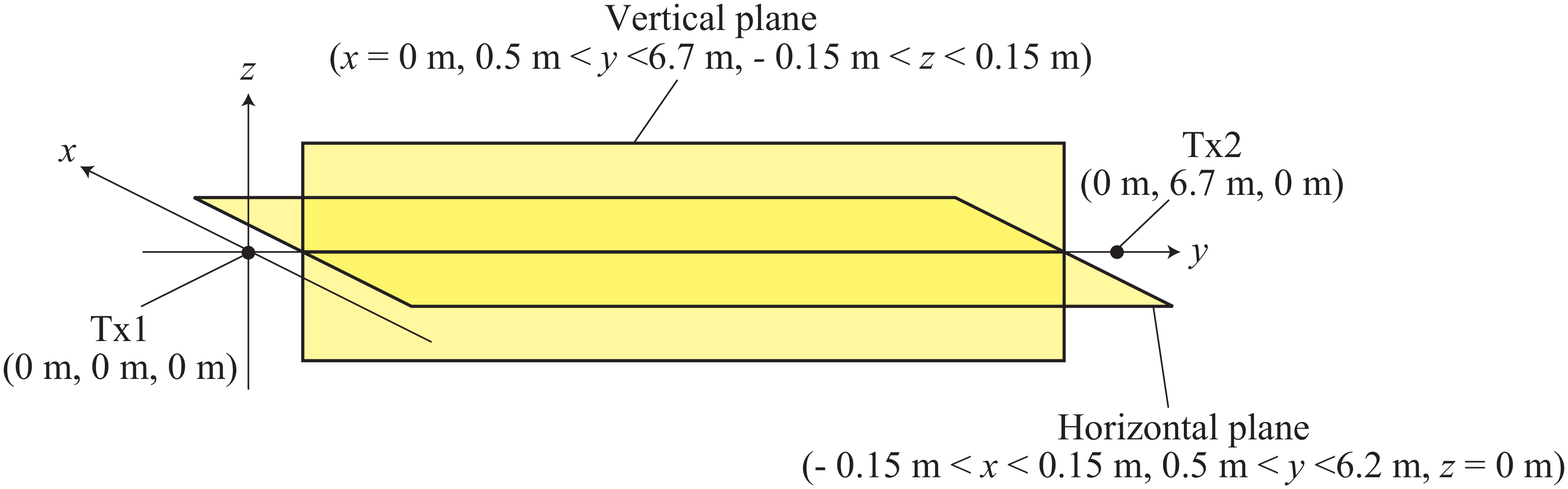} 
\caption{Measurement field.}
\label{fig:Meas} 
\end{figure}

\begin{figure}  [h]
\centering
\includegraphics[width=12cm]{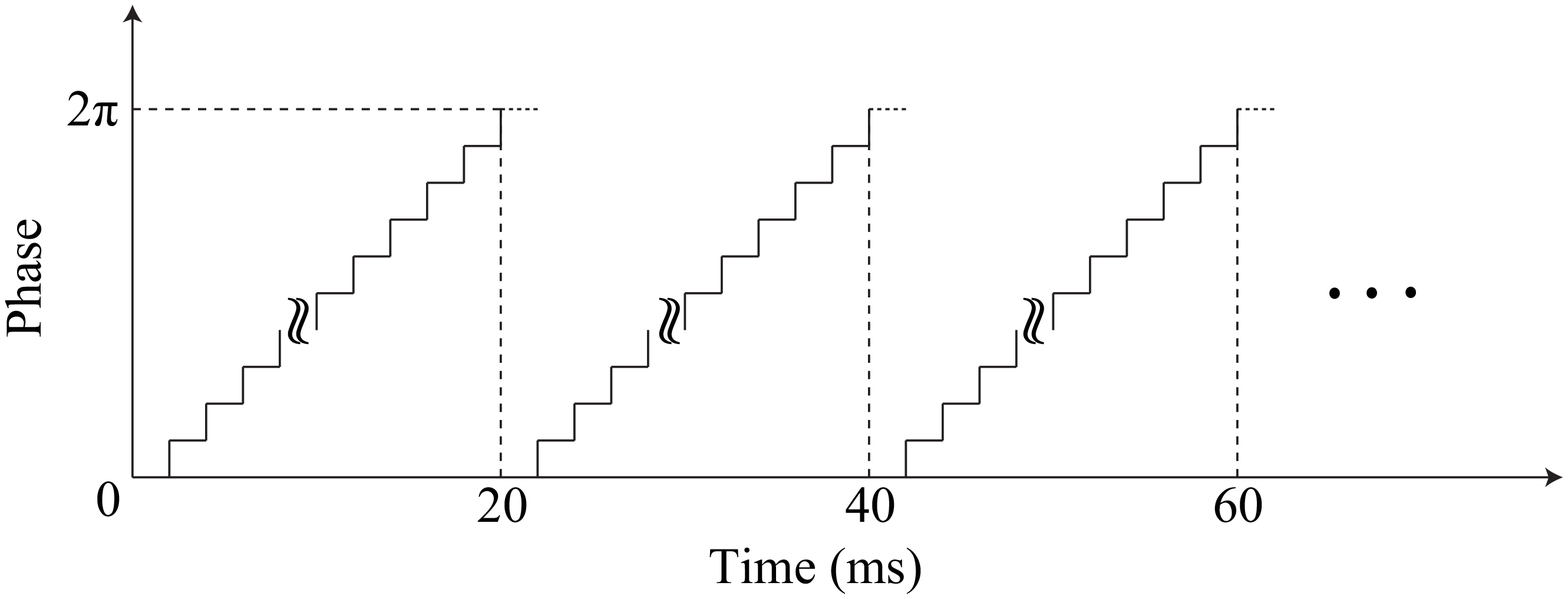} 
\caption{Relationship between phase and time in phase shifter.}
\label{fig:v_t} 
\end{figure}

The experimental system, environment and equipments are shown in Figs.~\ref{fig:Ex_model},~\ref{fig:Ex_model_pic} and~\ref{fig:ins} respectively. A RFID R/W (Fig.~\ref{fig:ins}(a)) equipped with horizontal patch antennas (Fig.~\ref{fig:ins}(b)) of 6 dBi gain is employed to perform wireless energy transmission,  while an IC tag antenna (Fig.~\ref{fig:ins}(c)) is used to receive energy. In addition, a variable phase shifter (Fig.~\ref{fig:ins}(d)) is used to produce carrier offset. The center frequency is 952.4 MHz and the transmit power per each antenna is 30 dBm. Both the Tx antenna $\sharp$1 and $\sharp$2 are set at the same height of 1.05 m. The measurement is performed in a three dimensional space. While the coordinates of Tx1 and Tx2 are fixed as (0 m, 0 m, 0 m) and (0 m, 6.7 m, 0 m), that of Rx is moved within the horizontal plane ($-0.15$ m $\leq x \leq$ $0.15$ m, $0.5$ m $\leq y \leq$ 6.2 m, $z$ = 0 m) and the vertical plane ($x =$ 0 m, 0.5 m $\leq y \leq$ 6.2 m, $-0.15$ m $\leq z \leq$ 0.15 m) by a positioner with a step of 3 cm to evaluate the coverage of wireless energy transmission as shown in Fig.~\ref{fig:Meas}. For SP, we consider both cases that wireless power are transmitted from either Tx antenna $\sharp$1 or $\sharp$2. For MP, signal from the RFID R/W is divided by a power divider, and then transmitted from both the Tx antenna $\sharp$1 and $\sharp$2. In the case of MPCSD, the carrier offset is produced by a variable phase shifter between the divider and the Tx antenna $\sharp$2. The output phase can be changed by imposing different voltages controlled by a D/A board equipped on a PC. In this experiment, continuous constant phase change is created by setting the phase to repeatedly increase from $0$ to $2\pi$ in a step of $2\pi/100$ as shown in Fig.~\ref{fig:v_t}. By this method, a carrier offset of 50 Hz apart from the center frequency 952.4 MHz can be generated.

\begin{table} [!t]
\renewcommand{\arraystretch}{1.3}
\caption{Experimental parameters.}
\centering
\begin{tabular}{l || c}
\hline			
  Parameter & Value \\
\hline			
Transmit power per each antenna & 30 dBm \\ 
Center frequency & 952.4 MHz\\
Carrier shift & 50 Hz\\
Measurement point interval & 3 cm$\approx \lambda/10$ \\ 
 \hline  
\end{tabular}
\end{table}

\subsection{Free Space Simulation}
In order to easily understand power distribution property and coverage performance of the three schemes, we conduct a simulation assuming a free space environment where the placement of Txs and Rx are the same as those of the experiments. 

\begin{figure} [!t]
\centering
\includegraphics[width=12cm]{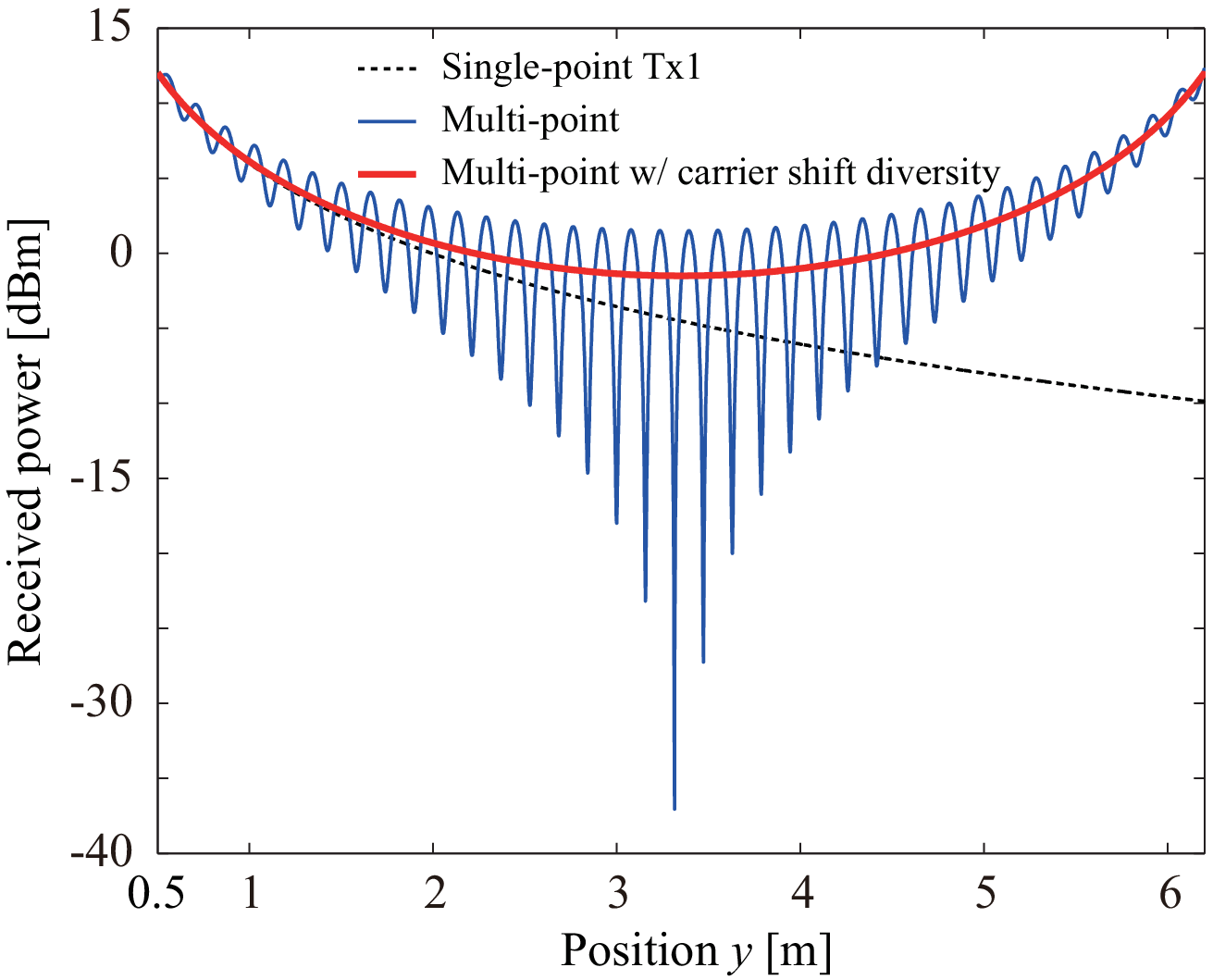} 
\caption{Power distribution in free space.}
\label{fig:PD_FS} 
\end{figure}

Figure~\ref{fig:PD_FS} shows the power distribution of each scheme along the straight line ($x =$ 0 m, 0.5 m $\leq y \leq$ 6.7 m, $z=$ 0 m). In SP, when the Rx antenna is far from the corresponding Tx antenna, the received power attenuates in proportion to free space path-loss. In MP, the area of energy supply field can be enhanced owing to the increased number of transmission points, as compared to that of SP. However, at the central area between the two Tx antennas, the received power is degraded by the effect of destructive interference between multiple wave sources. On the contrary, in MPCSD, the degraded received power at deadspots are remarkably improved as shown in Fig.~\ref{fig:PD_FS}.

\begin{figure} [!t]
\centering
\includegraphics[width=12cm]{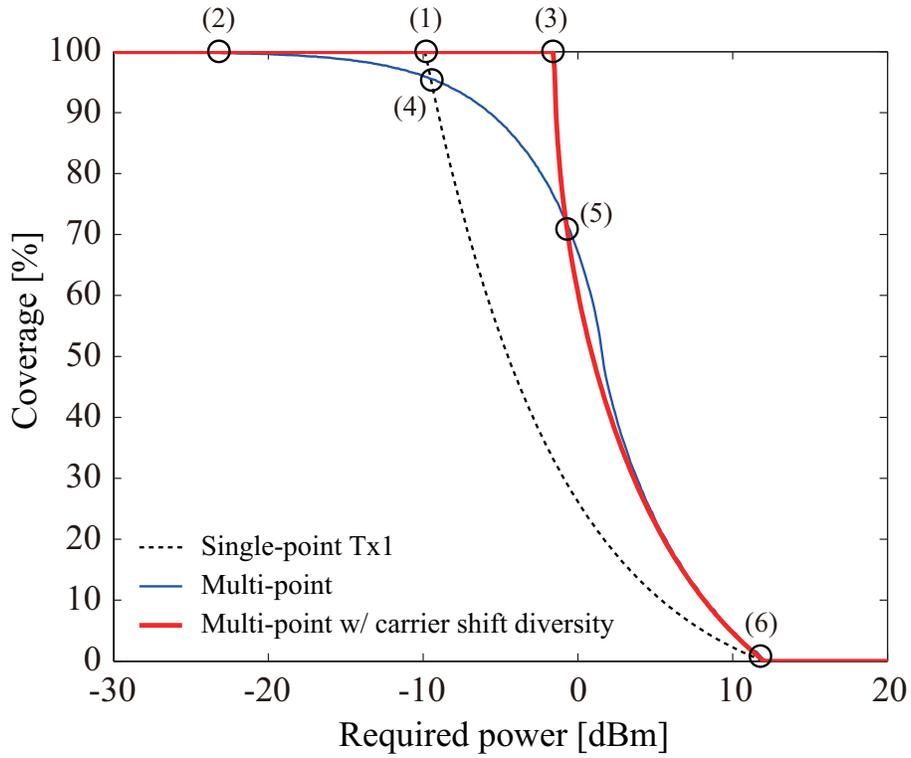} 
\caption{Energy transmission coverage in free space simulation.}
\label{fig:Co_FS} 
\end{figure}

 \begin{figure*} [!t]
\centering
\includegraphics[width=12cm]{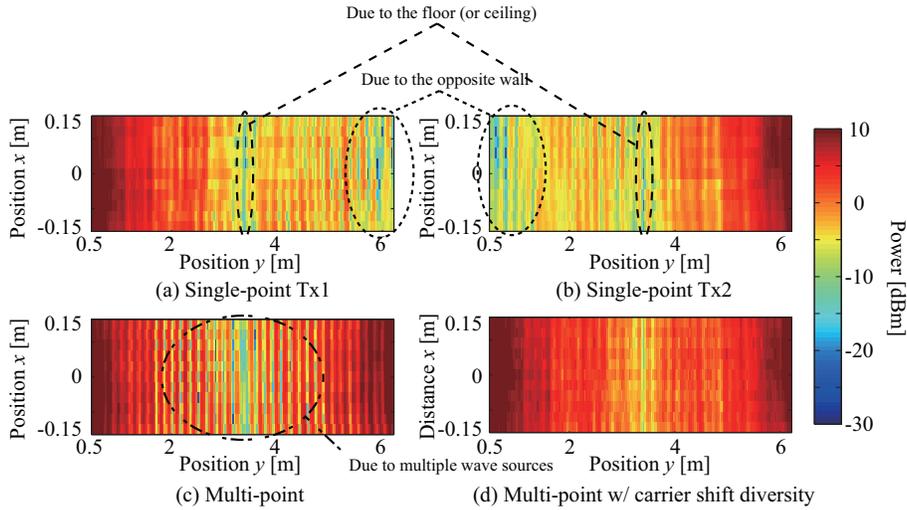} 
\caption{Power distribution in horizontal plane.}
\label{fig:PD_H} 
\end{figure*}

\begin{figure*} [!t]
\centering
\includegraphics[width=12cm]{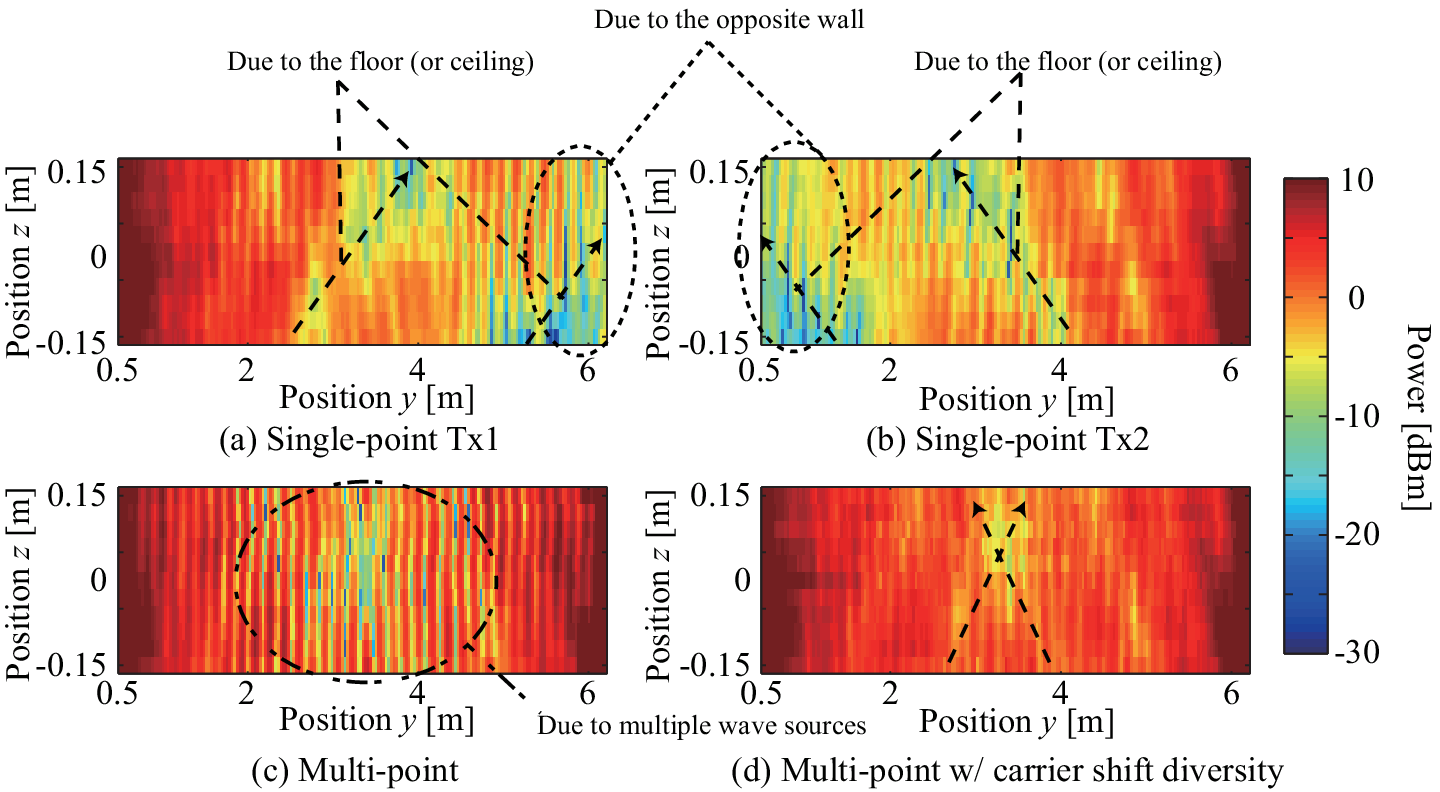} 
\caption{Power distribution in vertical plane.}
\label{fig:PD_V} 
\end{figure*}

\subsection{Definition of Coverage}
In order to evaluate the energy transmission schemes, the following metric is introduced. Equation (\ref{equ:coverage}) shows our definition of  coverage for wireless energy transmission.  

\begin{eqnarray}
{C}({{P}_{\mathrm{req}}})=\frac{\sum_{k=1}^{K}S(k, P_{\mathrm{req}})}{K}, 
\label{equ:coverage}
\end{eqnarray}
where $k$ denotes a measurement point,  $K$ is the total number of measurement points,  ${P}_{\mathrm{req}}$ denotes the required power to activate the sensor IC,  and ${S}(k, {P}_{\mathrm{req}})$ is the sensor activation indicator which is defined as
\begin{eqnarray}
S(k, P_{\mathrm{req}})=\left\{ \begin{array}{ll}
1 & (P_{\mathrm{r}}({k})\ge P_\mathrm{req}) \\
0 & (P_{\mathrm{r}}({k})<P_\mathrm{req}) \\
\end{array} \right.
,
\end{eqnarray}
where $P_{\mathrm{r}}({k})$ is the received power simulated or measured at point $k$. Therefore, the coverage expresses the ratio of area where the received power is higher than the required power $P_{\mathrm{req}}$ in the simulated or measured environment. 

Figure~\ref{fig:Co_FS} shows results of the coverage in the free space simulation. To evaluate each scheme, six special points are marked in the figure. Points (1) - (3) show the maximum available values of required powers to maintain 100\% coverage in SP, MP, and MPCSD respectively. Points (4) - (6) show the crossing points among SP, MP, and MPCSD. The maximum available value in SP as seen from the point (1) is determined by transmit power, path-loss attenuation, and antenna gain. Since MP creates deadspots due to destructive interference between multiple wave sources, the maximum available value in MP is much less than that in SP and MPCSD as seen from the points (1) - (3). On the other hand, the maximum available value of MPCSD is 8.4 dB higher than that in SP as shown by the gap between point (1) and (3).  In addition, the points (4) and (5) indicate that MP is only effective when supplying the power to certain sensors. Furthermore, all schemes converge to 0\% at the same required power as shown from the point (6). Therefore, Fig.~\ref{fig:Co_FS} shows that MPCSD is the most effective scheme to improve the uniformity of the coverage.

\subsection{Experiment Results}  
Measurement results on power distributions of SP in horizontal plane and vertical plane are shown in Figs.~\ref{fig:PD_H} (a)(b) and~\ref{fig:PD_V} (a)(b) respectively. When the Rx antenna is far from the corresponding Tx antenna, the received power attenuates in proportion to free space path-loss in both cases. In addition, when the Rx antenna is close to the opposite wall of the Tx antenna, the received power is fluctuated due to the standing-wave effect caused by reflection from the wall. Furthermore, at the measurement point around 3.5 m in the both cases of Tx antenna $\sharp$1 and Tx antenna $\sharp$2, the received power degrades due to the standing-wave caused by the reflections from the floor (or ceiling) in horizontal plane. In vertical plane, the degraded point changes depending on Rx antenna height $z$, because of the phase difference between the direct wave and the reflected wave from the floor (or ceiling). The results show that the coverage is limited by path-loss attenuation and deadspots due to the standing-wave created by multipath.

The power distributions of MP in horizontal plane and vertical plane are shown in Figs.~\ref{fig:PD_H} (c) and~\ref{fig:PD_V} (c) respectively. In MP, the area of energy supply field can be enhanced owing to the increased number of transmission points, as compared to SP in Fig.~\ref{fig:PD_H} (a)(b). In addition, since another Tx is additionally located at the opposite wall, the degradation due to the standing-wave caused by reflection from the wall can be reduced. However at the central area between the two Tx antennas, the received power is severely degraded by destructive interference between the two wave sources. On the contrary, in the case of MPCSD, the degraded received power at deadspots are remarkably improved as shown in Figs.~\ref{fig:PD_H} (d) and~\ref{fig:PD_V} (d). In terms of the deadspots caused by multipath, the number of deadspots decreases compared with that of SP since deadspots of MPCSD occur only when both the received powers from the two single-points are degraded at the same time. 

\subsubsection{Coverage of Wireless Energy Transmission}
\begin{figure} [t]
\centering
\includegraphics[width=12cm]{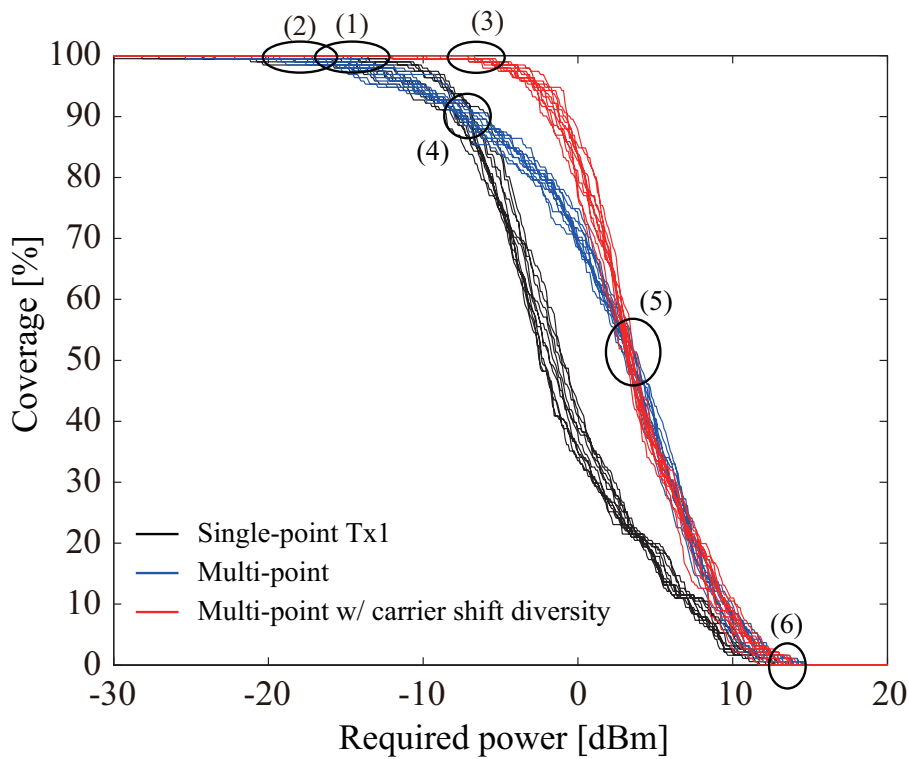} 
\caption{Experiment result of energy transmission coverage in horizontal plane.}
\label{fig:Co_H} 
\end{figure}

Figure~\ref{fig:Co_H} shows the coverage of each scheme measured in horizontal plane ($x$ - $y$ plane) at discrete $x =$ $\{$ $-0.15$, $-0.12$, $\ldots$, $0.15 $ m~$\}$. The trend of all schemes are similar to that of free space simulation in Fig.~\ref{fig:Co_FS}. However, as seen from the points (1) and (3) in the figure, the coverage in SP and MPCSD is shrunken compared to that of free space simulation while the coverage in MP is almost the same value as seen from the point (2). It implies that the effect of standing-wave caused by multipath and interference between multiple wave sources decreases the coverage performances.  

\begin{figure} [t]
\centering
\includegraphics[width=12cm]{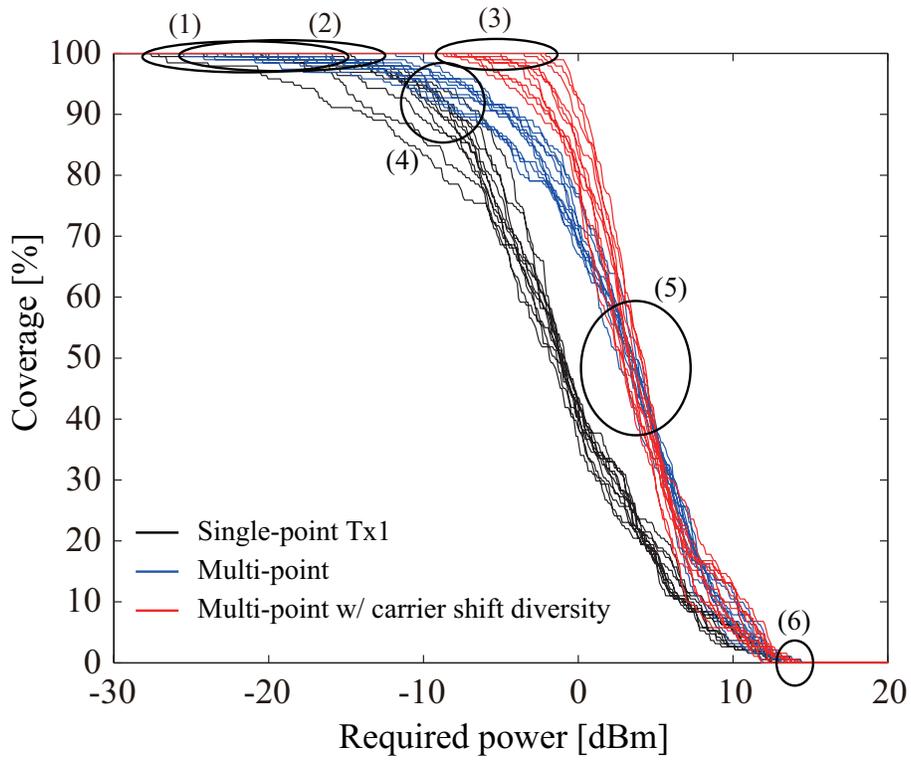} 
\caption{Experiment result of energy transmission coverage in vertical plane.}
\label{fig:Co_V} 
\end{figure}

Figure~\ref{fig:Co_V} shows the coverage of each scheme measured in vertical plane ($z$ - $y$ plane) at discrete $z =$ $\{$ $-0.15$, $-0.12$, $\ldots$, $0.15$ m~$\}$. The trend of all schemes are similar to the results in the case of free space simulation in Fig.~\ref{fig:Co_FS}. However, the maximum available values of required powers in SP and MPCSD vary with respect to $z$ more remarkably as compared to the horizontal case, as seen at the points (1) and (3) in Fig.~\ref{fig:Co_V}. It implies that power degradation due to reflections from the floor (or ceiling) depends on the Rx antenna height $z$. On the other hand, the maximum available values of required power in MP is similar to that in the horizontal case since the effect of interference between multiple wave sources is dominant compared to the effect of standing-wave caused by multipath, as seen from the point (2) in Fig.~\ref{fig:Co_V}.

\begin{figure} [t]
\centering
\includegraphics[width=12cm]{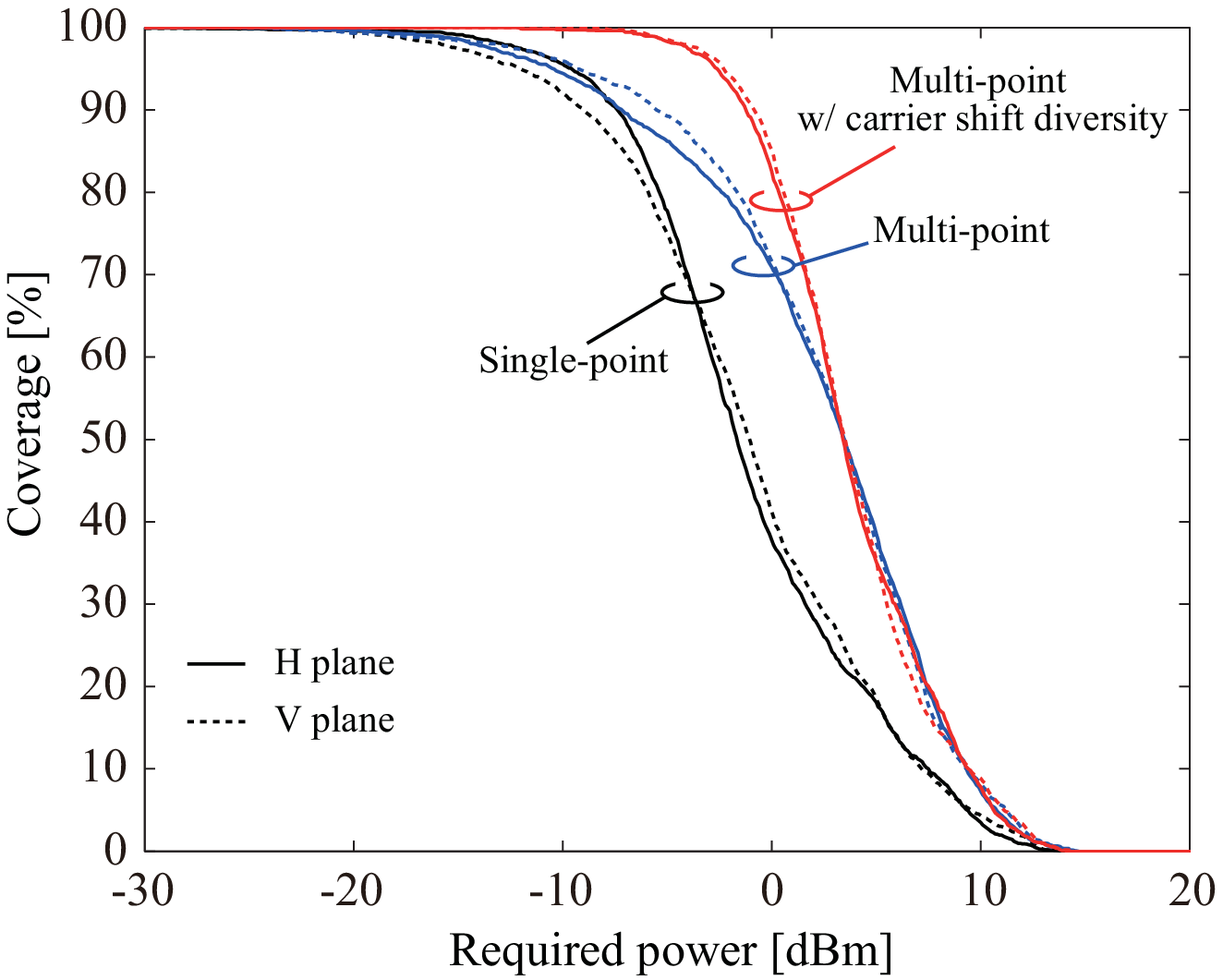} 
\caption{Experiment result of energy transmission coverage (overall performance).}
\label{fig:Co_all} 
\end{figure}

Finally, Fig.~\ref{fig:Co_all} shows the coverage of each scheme measured in the whole horizontal or vertical plane. In SP, the maximum available value of required power  in vertical plane is lower that that in horizontal plane. It means that the effect of multipath is most dominant for SP. In MP, the values of the two cases are almost the same. It is because that the effect of interference between multiple wave sources is dominant compared to that of standing-wave created by multipath in MP. In MPCSD, the values of required power in both horizontal and vertical case are also almost the same. It implies that the effect of standing-wave created by multipath can be reduced by MPCSD.  In horizontal and vertical planes, the values in MPCSD are respectively 18.2 dB and 25.6 dB higher than that in SP as shown in the figure, so that the gain of MPCSD in this experiment is much higher than that in the free space simulation.

\section{Conclusion}
\label{sec:4}
This paper conducted indoor experiments to verify the effectiveness of the multi-point wireless energy transmission with carrier shift diversity which can improve the coverage of energy supply field. We compared the received power distribution and the coverage performance of different energy transmission schemes including conventional single-point, simple multi-point and our proposed multi-point scheme. To easily observe the effect of standing-wave caused by multipath and interference between multiple wave sources, the measurements were performed in a three dimensional space of an empty room and also simulated in free-space conditions. The experimental results showed that standing-wave due to multipath and interference between multiple-wave sources are respectively dominant in the single-point scheme and in the simple multi-point scheme. On the other hand, in the proposed multi-point scheme, the effect of standing-wave created by multipath and interference between multiple wave sources can be mitigated. In this experimental environment, the maximum available values of required power in the proposed scheme in horizontal and vertical planes are respectively 18.2 dB and 25.6 dB higher than that of the single-point scheme while the gain was 8.4 dB in free space simulation. It can be concluded that the proposed scheme can mitigate power attenuation due to the path-loss as well as the effect of standing-wave created by multipath and interference between multiple wave sources, so that the proposed scheme can improve the coverage of energy supply field.

For future works to improve the uniformity of the coverage of energy supply field, we need to consider the design of antenna directivity, antenna polarization and density of transmitters at given conditions of target environments.

\end{document}